\documentclass[12pt,preprint]{aastex}

\slugcomment{to be published in the {\it Astrophysical Journal {\it Letters}}}
\shorttitle{{\it Chandra} Limits of Supermassive Black Holes}
\shortauthors{Loewenstein et al.}

\begin{document}

\title{{\it Chandra} Limits on X-ray Emission Associated with the
Supermassive Black Holes in Three Giant Elliptical Galaxies}
\author{Michael Loewenstein\altaffilmark{1}, Richard F. Mushotzky,
Lorella Angelini\altaffilmark{2}, Keith A. Arnaud\altaffilmark{1}}
\affil{Laboratory for High Energy Astrophysics, NASA/GSFC, Code 662,
Greenbelt, MD 20771}
\email{loew@larmes.gsfc.nasa.gov}

\and

\author{Eliot Quataert}
\affil{Institute for Advanced Study, School of Natural Sciences,
Einstein Drive, Princeton, NJ 08540}
\email{eliot@ias.edu}
\altaffiltext{1}{Also with the University of Maryland Department of
Astronomy}
\altaffiltext{2}{Also with the Universities Space Research
Association}

%===============================================================================
\begin{abstract}
Elliptical galaxy nuclei are the sites of the largest black holes
known, but typically show little or no nuclear activity.  We
investigate this extreme quiescence using {\it Chandra X-ray
Observatory} observations of the giant elliptical galaxies NGC 1399,
NGC 4472, and NGC 4636. The unique {\it Chandra} imaging power enables
us to place upper limits of 7.3, 15 and $28\times 10^{-9}$ the
Eddington luminosity for the $\sim 10^8-10^9$ M$_{\odot}$ black holes
in NGC 1399, NGC 4472, and NGC 4636, respectively. The corresponding
radiative efficiencies in this band are 4.1, 24, and $620\times
10^{-6}$ using Bondi accretion rates derived from the {\it Chandra}
hot interstellar gas surface brightness profiles.  These limits are
inconsistent with basic advection-dominated accretion flow (ADAF)
models for NGC 1399 and NGC 4472, indicating accretion onto the black
hole at $\lesssim 10$\% of the Bondi rate.
\end{abstract}


%===============================================================================

\section{Introduction}

Knowledge about the frequency and masses of supermassive black holes
(SMBH) in galactic nuclei, and the characteristics of their associated
radiation, deepen our understanding of galaxy formation and clarify
its connection with the active galactic nuclei (AGN) phenomenon (e.g.,
Blandford 1999). While the intense energetics and rapid variability in
the most luminous active galaxies form the empirical foundation of the
SMBH/AGN paradigm, such extreme objects are rare -- especially in the
local universe.

Nevertheless, there is now a consensus that most, if not all, galaxies
host SMBH. This is based, in part, on the high frequency of low-level
nuclear activity in nearby galaxies \citep{HFS}, and on statistical
arguments connected to the cosmic X-ray background (e.g., Fabian \&
Iwasawa 1999).  However, the strongest evidence for the ubiquity of
SMBH comes from dynamical studies. \citet{Ma98} found that, of 32
galaxies that could be fit with an axisymmetric dynamical model, 26
(29) require the presence of a central massive ($\sim 10^8-10^{10}$
M$_{\odot}$) dark object at 95\% (68\%) confidence.  More unambiguous
and accurate measurements based on {\it Hubble Space Telescope} ({\it
HST}) stellar or ionized gas spectroscopy confirms the presence of
SMBH while deriving systematically lower SMBH masses
\citep{Ge00a,MF01a}.

Although kinematically-studied luminous AGN are found to host SMBH
\citep{Ge00b,MD00}, the converse is not true: activity in most SMBH
host galaxies is modest or absent. This can be quantified by
considering the ratio of the nuclear luminosity to the Eddington
luminosity ($L_{\rm edd}=1.25\ 10^{38}M_{\rm SMBH}$, where the black
hole mass $M_{\rm SMBH}$ is measured in solar units). The most powerful
AGN have ratios on the order of unity, as expected from the theory of
thin disks accreting at their maximal rate.  However, this ratio is
less than $10^{-7}$ and $10^{-4}$, respectively, in the two most
securely measured SMBH -- SgrA$^*$ at the Galactic Center \citep{QNR}
and NGC 4258 \citep{GNB}. If these SMBH accrete at the \citet{Bo52}
rate, they must do so in a fundamentally different -- and much less
radiatively efficient -- way than in luminous AGN. The theory of
advection-dominated accretion flows (ADAF; e.g., Narayan \& Yi 1994),
provides a predictive formalism for explaining the observed quiescence
in the nuclei of SgrA$^*$, NGC 4258, and other systems
\citep{FR95,Ma97}.

Elliptical galaxies are the sites of the most luminous quasars, and
also host the most extreme cases of quiescent SMBH. Although they
contain nuclear black holes with the highest estimated masses and are
observed to contain an ample supply of fuel in the form of X-ray
emitting hot interstellar gas, they often emit less than
$10^{-6}L_{\rm edd}$ \citep{Lo98,Di00}. Radio/X-ray observations of
some systems are inconsistent with predictions of the simplest models
of an ADAF accreting at the Bondi rate, but can be accommodated by
variations that include outflow \citep{BB99,QN99,Di99} or convection
\citep{NIA,QG00}.

Prior to the launch of the {\it Chandra X-ray Observatory}, X-ray
searches for nuclear point sources in elliptical galaxies were
complicated in the imaging domain by centrally concentrated hot gas
that dominates the soft X-ray flux; and, in the spectral domain by an
extended hard component consisting of X-ray binaries. Thus the
previous sensitivity to SMBH-associated X-ray emission, for even the
nearest giant ellipticals, was $\sim 10^{40}$ erg s$^{-1}$
\citep{RW00,SB01}, as neither the {\it ASCA} SIS/GIS nor the {\it
ROSAT} PSPC/HRI had the necessary combination of sensitivity and
spatial and spectral resolution to probe more deeply.  \citet{ADF}
claim to detect a very hard nuclear X-ray component in six elliptical
galaxies with characteristics consistent with ADAF/outflow models,
given the observed radio emission \citep{Di99}.  However, these
results are based on a non-unique decomposition of {\it ASCA} spectra.

Because of the presence of bright stellar cores and dust, the best
approach for constraining the level of SMBH activity is via high
angular resolution X-ray observations (e.g., George et al. 2001): for
a typical AGN X-ray/optical ratio, the {\it Chandra} detection
threshold corresponds to a B magnitude $\sim 24$, while elliptical
galaxy central surface brightnesses range from $\sim 11-16$ mag
arcsec$^{-2}$. This is further demonstrated by deep {\it Chandra}
surveys in combination with optical follow-up.  Low-luminosity AGN,
most not optically detected, dominate the X-ray background; and,
nonthermal emission is found to be common -- at any given time $\sim
10$\% of all bright galaxies host $L_X\gtrsim 10^{41}$ erg s$^{-1}$
AGN \citep{Mu00,Bar01}. There is also a high frequency of nuclear
X-ray point sources associated with local AGN, though often at levels
that overlap in luminosity with X-ray binaries \citep{Ho01}.  {\it
Chandra} has detected unresolved nuclear emission associated with
quiescent SMBH in M31 at $\sim 5\times 10^{-9}L_{\rm edd}$
\citep{Ga00}, the Galactic Center at $\sim 10^{-11}L_{\rm edd}$
\citep{Bag01}, and NGC 6166 at $\sim 10^{-7}L_{\rm edd}$ \citep{Di01}.

In this paper we use {\it Chandra} imaging data to place luminosity
limits on any nuclear point sources in the elliptical galaxies NGC
1399, NGC 4472, and NGC 4636. These limits raise new questions about
the radiative efficiency and rate of accretion onto supermassive black
holes.  Following \citet{MF01b}, we adopt distances of 20.5, 16.7, and
15.0 Mpc for NGC 1399, NGC 4472, and NGC 4636, respectively.

%===============================================================================

\section{Data Analysis and Results}

\subsection{Observations, Data Reduction and Analysis}

NGC 1399, NGC 4472, and NGC 4636 were observed with the {\it Chandra}
ACIS-S detector for 40--60 ksec during Cycle 1. Results on the diffuse
hot gas emission and the X-ray binary population, and a more detailed
data reduction description, are included elsewhere (e.g, Angelini,
Loewenstein, \& Mushotzky 2001).  The latest revised standard pipeline
processing was used for NGC 4472 and NGC 4636, while an earlier
version in combination with additional bad pixel removal, gain map
correction, and quantum efficiency map correction was used for NGC
1399. Final useful exposure times are 56, 40, and 52 ksec for NGC
1399, NGC 4472, and NGC 4636, respectively.  NGC 1399 was
approximately centered on the S3 chip; NGC 4636 and NGC 4472 were
positioned at the standard ACIS-S aimpoint.  1.25--7 keV images of the
central $50''\times 50''$ are shown for all three galaxies in Figures
1a, 1b, and 1c. The positions of the optical nuclei from archival {\it
HST} NICMOS (NGC 4472, NGC 4636) or WFPC2 (NGC 1399) images are
marked.  The brightest point sources in the images are easily visible
and indicated in the images; their counting rates are in the
0.005--0.008 cts s$^{-1}$ range. The bright X-ray core in NGC 4472 is
resolved and has a thermal spectrum.  Only in the case of NGC 4636 is
there a possible coincidence of the nucleus with a {\it Chandra} X-ray
source (just to the southwest of the nuclear position indicated in
Figure 1c), with a 0.3--7 keV count rate $\sim 0.004$ cts s$^{-1}$
(see below). 0.3--7.0 keV surface brightness profiles are shown in
Figure 2. For illustration purposes, the NGC 4636 profile is centered
on the near-nuclear source.  No excess emission from unresolved
nuclear emission is apparent for the other two galaxies.

The position of the nucleus in NGC 1399 is identified based on
registration of globular cluster X-ray sources with their WFPC2
counterparts (Angelini et al. 2001); the resulting {\it HST} and {\it
Chandra} centroids are consistent. Similarly, the nucleus of NGC 4636
is located based on registration with a bright NICMOS source and with
the USNO catalog \citep {M98}, and is found {\it not} to be consistent
with the position of the near-nuclear source (Figure 1c).  For NGC
4472, we adopt the unregistered {\it HST} centroid that is consistent
with the peak in the soft thermal emission seen in the broadband {\it
Chandra} image (Figure 1b).  $3\sigma$ upper limits to any nuclear
X-ray point source emission are derived following \citet{Ge86}. Counts
are accumulated in inner (``source'') and outer (``background'')
concentric annuli with radii that are varied to obtain the tightest
upper limit.  The {\it Chandra} calibration database enclosed energy
function at the appropriate off-axis angle is used, and an exposure
map correction is applied for NGC 1399, which was centered off-axis.
The assumed position of the SMBH is allowed to vary in a $3''\times
3''$ box (corresponding to a very conservative adopted {\it
Chandra}/{\it HST} relative systematic pointing uncertainty) centered
on the assumed nuclear positions and the highest upper limit in this
box is adopted. These boxes encompass the peaks of the diffuse
emission for NGC 4472 and NGC 1399.  The diffuse hot gas emission is
implicitly assumed to be locally flat on average when treated as
``background'', consistent with the flat central surface brightness
profiles (Figure 2).  Luminosity upper limits are derived both in the
0.3--7 keV and 2--10 keV energy bands assuming a slope 1.5 power-law
spectrum -- the limits increase by $\sim 2$ for slope 0.5. Due to
reduced dilution from the diffuse hot gas emission, the 2--10 keV
count rate limits are 3--5 times lower than in the softer band;
however, because of the declining effective area the energy flux upper
limits are comparable. The 0.3--7 keV detection limit varies from a
few $10^{37}$ erg s$^{-1}$ in regions of low surface brightness hot
ISM emission to a few $10^{38}$ erg s$^{-1}$ in the core of NGC 1399.

The 2--10 keV limits -- all derived within $1''$ of the optical
nucleus -- are displayed in Table 1; these correspond to 28, 23, and
16 counts for NGC 1399, NGC 4472, and NGC 4636, respectively --
comparable to the brightest sources in the subimages shown in Figure
1. Table 1 also includes less conservative, but perhaps more
realistic, limits calculated by fixing the positions at the assumed
nuclear positions described above.

The inner gas density profile is estimated by fitting the
point-source-subtracted surface brightness profiles with
$\beta$-models,
\begin{equation}
\Sigma=\Sigma_o\left(1+{{r^2}\over {a^2}}\right)^{-(3\beta-1/2)}
\end{equation}
out to some maximum radius, followed by analytical deprojection with
conversions obtained from spectral fits to the inner regions (best-fit
temperatures $kT=0.8$, 0.8, and 0.6 keV for NGC 1399, NGC 4472, and
NGC 4636, respectively). The quality of fit declines as the maximum
radius is increased due to deviations from the $\beta$-model, but the
profiles are sufficiently concentrated that the central density
estimates -- 4.6, 3.2, and $1.1\times 10^{-25}$ gm cm$^{-3}$ for NGC
1399, NGC 4472, and NGC 4636, respectively -- are robust.

\subsection{Accretion Rates and Eddington Ratio Limits}

Individual black hole masses for the three galaxies studied here have
been estimated by \citet{Ma98} from two-integral dynamical modeling of
ground-based kinematical data.  A more accurate value can be derived
using the correlation with bulge velocity dispersion, $\sigma$,
derived for galaxies with limits based on {\it HST}
spectroscopy. Therefore, we adopt SMBH masses from \citet{MF01b}
(Table 1); these are 3--5 times lower than in \citet{Ma98}. The
resulting 2--10 keV luminosities in units of the Eddington luminosity
and the \citet{Bo52} accretion rates for adiabatic spherical inflow
are displayed in Table 1.  Our derived upper limits on $L_X/L_{\rm
edd}$ are $7.3\times 10^{-9}$, $1.5\times 10^{-8}$, and $2.8\times
10^{-8}$ for NGC 1399, NGC 4472, and NGC 4636, respectively. The
corresponding efficiencies, assuming accretion at the Bondi rate $\dot
M_{\rm Bondi}$, are $4.1\times 10^{-6}$, $2.4\times 10^{-5}$, and
$6.2\times 10^{-4}$.

%===============================================================================

\section{The Nature of the Accretion Flow}

The conservative upper limits we derive are 1--2 orders of magnitude
lower than the luminosities estimated by Allen et al. (2000) using
{\it ASCA} spectral decomposition. Direct {\it Chandra} images
(Angelini et al. 2001) indicate that most of the hard X-ray emission
spatially unresolved by {\it ASCA} originates in X-ray binaries.
Since upper limits on the nuclear X-ray emission and estimates of the
Bondi accretion rate have each decreased by $\sim 10-30$ from
\citet{Di00} (the latter because of lower black hole mass
estimates)\footnote{For NGC 4636 our Bondi accretion rate is $\approx
3000$ times smaller than that of \citet{Di00}.  A factor of $\approx
10$ is due to our smaller black hole mass; the remainder is primarily
due to a lower inferred central gas density that may be underestimated
since the accretion radius is unresolved in this case.}, a
re-examination of the nature of the accretion flow is in order.

For our estimated black hole masses and Bondi accretion rates, we have
computed spectra using the techniques described in \citet{QN99}; we
aim to test the basic ADAF model and so do not include outflows or
convection. The resulting predicted 2--10 keV luminosities, $L_{\rm
ADAF}$, are $\approx 2 \times 10^{41}$, $10^{40}$, and $10^{36}$ ergs
s$^{-1}$ for NGC 1399, NGC 4472, and NGC 4636, respectively (Table
1). Our observational upper limits are thus significantly below the
ADAF predictions for NGC 1399 and NGC 4472, but consistent for NGC
4636. The predicted X-ray luminosities in the standard ADAF models for
the cases of NGC 1399 and NGC 4472 are dominated by Compton scattered
synchrotron emission that originates close to the black hole. The
luminosity is primarily sensitive to the gas density (i.e., accretion
rate) and electron temperature.  A factor of $\sim 10$ decrease in the
accretion rate below the Bondi value can reduce the predicted
luminosity to the observational limits.

With the revised black hole masses and accretion rates we find that
the radio flux predicted by the ADAF model is less than the observed
flux for NGC 1399 and NGC 4636, but is in excess for NGC 4472; thus
the radio flux problem highlighted by \citet{Di99} is somewhat
ameliorated.  Finally, a preliminary {\it HST} measurement of the
nuclear FUV emission in NGC 1399 (0'Connell et al., in preparation),
-- $\nu L_{\nu}\sim 7\ 10^{38}$ erg s$^{-1}$ -- also conflicts with
pure-inflow ADAF models, although dust may be an issue at this
wavelength.

Our observations require a modification of the standard ADAF/Bondi
model such that the gas density near the SMBH is reduced by at least a
factor of $\sim 10$.  This could be due to the role of strong outflows
or convection in an ADAF.  Alternatively, the central regions of these
galaxies may not be in steady state due to heating by supernovae or by
episodic accretion onto the central SMBH (e.g., Ciotti \& Ostriker
1997).  In addition the cooling time at the accretion radius $R_{\rm
Bondi}$ (Table 1) is $<10^7$ yr for all three galaxies; so, inflow
with mass deposition could be important \citep{BM99}.

%===============================================================================
\section{Conclusions}

We have used {\it Chandra} observations to place new and unprecedented
upper limits on the X-ray emission associated with $\sim 10^8-10^9$
M$_{\odot}$ nuclear black holes in three giant elliptical
galaxies. The black holes clearly are not fuel-starved: all have
central hot gas densities exceeding $10^{-25}$ gm cm$^{-3}$.  Adopting
masses from the $M_{\rm SMBH}-\sigma$ correlation, and assuming that
mass is inflowing at the \citet{Bo52} rate, we derive $3\sigma$ 2--10
keV luminosity upper limits of 7.3, 15 and $28\times 10^{-9}L_{\rm
Edd}$ and radiative efficiencies in this band of 4.1, 24, and
$620\times 10^{-6}$ for NGC 1399, NGC 4472, and NGC 4636,
respectively.  Unless the black hole masses are considerably lower
than predicted by the $M_{\rm SMBH}-\sigma$ correlation, these
observations are inconsistent with the simplest ADAF models in NGC
1399 and NGC 4472, and imply accretion onto the SMBH at below the
Bondi rate.

If the decline in the accretion rate is due to outflow, this raises
the question of the ultimate fate of this gas as it encounters the
dense hot interstellar medium in the core.  Many questions about
stability, intermittency, and regulation of gas flows in the inner
regions of ellipticals remain to be addressed.  The {\it Chandra}
isophotes in the cores of NGC 1399 and NGC 4472 appear round and
undisturbed, although there is an apparent spiral-like structure in
the NGC 4636 diffuse emission.

The more stringent limits required to further constrain the nature of
the central accretion flow will be difficult. Our observations
indicate $>0.01$ super-Eddington X-ray binaries per square-arcsecond
near the nucleus; so that detections below $\sim 10^{39}$ erg s$^{-1}$
cannot be unambiguously identified as AGN.  Although the hot gas
emissivity precipitously declines at energies above 2 keV, so does the
{\it Chandra} sensitivity, restricting the detection limit in gas-rich
systems.  Tighter limits can be obtained on gas poor systems that,
although of interest, may be fuel-starved and would thus shed little
light on the accretion process.  It would be especially useful to
spectrally decompose the inner $\sim 1''$ to place a limit on the
fraction of non-ISM emission. This would appear as a hard excess and
would require a very deep exposure due to the declining {\it Chandra}
effective area mentioned above.

\acknowledgments

We are grateful to Dave Davis and Una Hwang for assistance with data
reduction, and to Wayne Landsman for sharing unpublished {\it HST}
results. We made use of the {\it HST} archive, the NASA/IPAC
Extragalactic Database (NED), and IDL function fitting routines
developed by Craig Markwardt.

%===============================================================================

\clearpage

%=============================================================================== 
\begin{deluxetable}{cccccccccccc}
%\tabletypesize{\scriptsize}
\tablecaption{Galaxy and Accretion Flow Characteristics}
\tablewidth{0pt}
\tablehead{
\colhead{galaxy} & \colhead{$d$} & 
\colhead{$M_{\rm SMBH}$} & 
\colhead{$R_{\rm Bondi}$} & 
\colhead{$\dot M_{\rm Bondi}$} & 
\colhead{$L_{\rm edd}$} & 
\colhead{$L_{\rm Bondi}$\tablenotemark{a}} &
\colhead{$L_{\rm ADAF}$\tablenotemark{b}} &
\colhead{$L_X$\tablenotemark{c}} &
\colhead{$L_X$\tablenotemark{d}} \\
NGC & Mpc & $10^8$ M$_{\odot}$ & $''$ & M$_{\odot}$ yr$^{-1}$ & erg
   s$^{-1}$ & erg s$^{-1}$ & erg s$^{-1}$ & erg s$^{-1}$ & erg
   s$^{-1}$}
\startdata
1399 & 20.5 & 10.6 & 0.36 & $4.0\ 10^{-2}$ & $1.3\ 10^{47}$ & $2.3\
10^{44}$ & $2\ 10^{41}$ & $<9.7\ 10^{38}$ & $<9.7\ 10^{38}$ \\
4472 & 16.7 & 5.65 & 0.24 & $7.9\ 10^{-3}$ & $7.1\ 10^{46}$ & $4.5\
10^{43}$ & $10^{40}$ & $<6.4\ 10^{38}$ & $<4.9\ 10^{38}$ \\
4636 & 15.0 & 0.791 & 0.049 & $8.0\ 10^{-5}$ & $9.9\ 10^{45}$ & $4.5\
10^{41}$ & $10^{36}$ & $<2.7\ 10^{38}$ & $<1.8\ 10^{38}$ \\
\enddata

\tablenotetext{a}{$0.1\dot M_{\rm Bondi}c^2$}
\tablenotetext{b}{approximate expectation of standard ADAF model; see text}
\tablenotetext{c}{2--10 keV upper limit from this paper for $3''$ box}
\tablenotetext{d}{2--10 keV upper limit at assumed optical nucleus}

\end{deluxetable}

\clearpage

%=============================================================================== 

\begin{figure}
\figurenum{1}
\centerline{\includegraphics[scale=0.70,angle=-90]{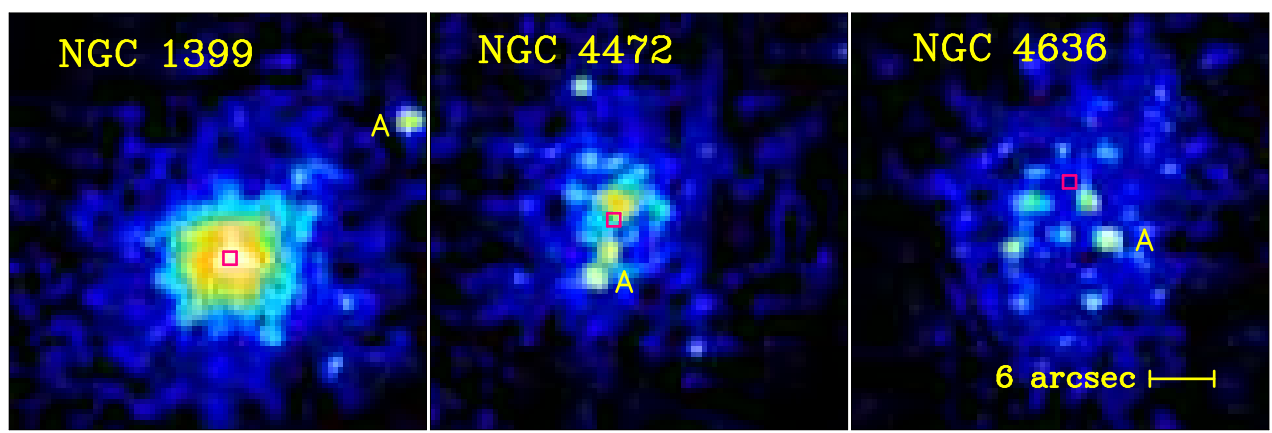}}
\caption{{\it Chandra} ACIS-S3 1.25--7 keV images of the central
$50''\times 50''$ of (a) NGC 1399, (b) NGC 4472, and (c) NGC 4636.  The
open, $1''\times 1''$ box indicates the position of the optical ({\it
HST}) nucleus, the letter ``A'' the brightest point source. Relative
shifts based on image registration have been applied for NGC 1399
($1.5''$)and NGC 4636 ($0.9''$).}
\end{figure}

\clearpage

\begin{figure}
\figurenum{2}
\centerline{\includegraphics[scale=0.70,angle=-90]{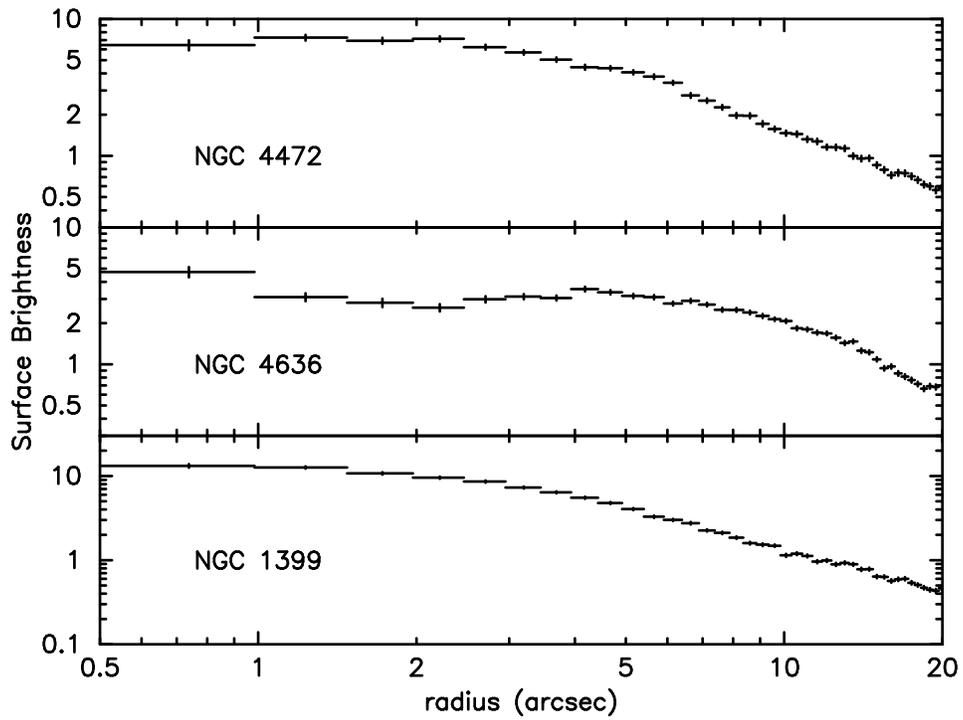}}
\caption{{\it Chandra} ACIS-S3 0.3--7 keV central diffuse emission
surface brightness distributions in cts arcmin$^{-2}$ s$^{-1}$.
Annuli are centered on the X-ray centroids for NGC 1399 and NGC 4472,
and on the near nuclear source for NGC 4636.}
\end{figure}

\clearpage

\end{document}